\documentstyle[12pt]{article}
\def\figcap{\section*{Figure Captions\markboth
        {FIGURECAPTIONS}{FIGURECAPTIONS}}\list
        {Figure \arabic{enumi}:\hfill}{\settowidth\labelwidth{Figure
999:}
        \leftmargin\labelwidth
        \advance\leftmargin\labelsep\usecounter{enumi}}}
 \relax
\newskip\humongous \humongous=0pt plus 1000pt minus 1000pt
\def\caja{\mathsurround=0pt}
\def\eqalign#1{\,\vcenter{\openup1\jot \caja
        \ialign{\strut \hfil$\displaystyle{##}$&$
        \displaystyle{{}##}$\hfil\crcr#1\crcr}}\,}
\newif\ifdtup

\def\ap#1#2#3{Ann.\ Phys.\ (NY) #1 (19#3) #2}

\def\jmp#1#2#3{J.\ Math.\ Phys.\ #1 (19#3) #2}

\def\np#1#2#3{Nucl.\ Phys.\ B#1 (19#3) #2}
\def\pl#1#2#3{Phys.\ Lett.\ #1B (19#3) #2}
\def\pr#1#2#3{Phys.\ Rev.\ D #1 (19#3) #2}
\def\prb#1#2#3{Phys.\ Rev.\ B #1 (19#3) #2}
\def\prep#1#2#3{Phys.\ Rep.\ #1 (19#3) #2}

\def\rmp#1#2#3{Rev.\ Mod.\ Phys.\ #1 (19#3) #2}

\def\cmp#1#2#3{Comm.\ Math.\ Phys.\ #1 (19#3) #2}
\def\cmp#1#2#3{Comm.\ Math.\ Phys.\ #1 (19#3) #2}
\def\nc#1#2#3{Il Nuovo Cimento #1A (19#3) #2}
\def\lnc#1#2#3{Lettere al Nuovo Cimento #1 (19#3) #2}
\def\apa#1#2#3{Acta Phy. Austriaca #1 (19#3) #2}
\newcounter{hran}
\def\bmini{\setcounter{hran}{\value{equation}}
\refstepcounter{hran} \setcounter{equation}{0}
\renewcommand{\theequation}{\thehran\alph{equation}}
              \begin{eqnarray}  }
\def\bminia{\setcounter{hran}{\value{equation}}
\refstepcounter{hran} \setcounter{equation}{1}
\renewcommand{\theequation}{\thehran\alph{equation}}
              \begin{eqnarray}  }
\def\bminiG#1{
          \setcounter{hran}{\value{equation}}
          \refstepcounter{hran}
          \setcounter{equation}{-1}
          \renewcommand{\theequation}{\thehran\alph{equation}}
          \refstepcounter{equation}
    \label{#1}
          \begin{eqnarray}          }
\def\emini{\end{eqnarray}\setcounter{equation}{\value{hran}}
\renewcommand{\theequation}{\arabic{equation}}}
\newskip\humongous \humongous=0pt plus 1000pt minus 1000pt
\def\caja{\mathsurround=0pt} \def\eqalign#1{\,\vcenter{\openup1\jot
\caja   \ialign{\strut \hfil$\displaystyle{##}$&$
\displaystyle{{}##}$\hfil\crcr#1\crcr}}\,} \newif\ifdtup

\def\half{\mbox{\small $\frac{1}{2}$}}

\def\ltap{\raisebox{-.4ex}{\rlap{$\sim$}} \raisebox{.4ex}{$<$}}

\def\frac#1#2{ {{#1} \over {#2} }}

\def\s#1{{\small #1}}
\def\fun#1#2{\lower3.6pt\vbox{\baselineskip0pt\lineskip.9pt
  \ialign{$\mathsurround=0pt#1\hfil##\hfil$\crcr#2\crcr\sim\crcr}}}
\def\ie{\hbox{\it i.e.}{ }}      
      
\def\partder#1{{\partial   \over\partial #1}}

\def\re#1{(\ref{#1})}
\def\beq{\begin{equation}}
\def\eeq{\end{equation}}
\def\beeq{\begin{eqnarray}}
\def\eeeq{\end{eqnarray}}
\def\bc{\bar c}
\def\bp{\bar p}

\def\r{\rho}
\def\s{\sigma}
\def\S{\Sigma}
\def\G{\Gamma}
\def\eps{\epsilon}
\def\bG{ \bar \Gamma}
\def\L{ \Lambda}
\def\l{ \lambda}
\def\g{ \gamma}

\def\d4#1{\frac {d^4 {#1} }{(2\pi)^4}}
\def\dL{\L \partial_\L }

\def\dim{n_A+n_c+2n_w+2n_v}
\def\UV{$\L_0\to\infty\;$}
\def\IR{$\L\to 0\;\;$}
\def\bit{\begin{itemize}}
\def\eit{\end{itemize}}
\def\ben{\begin{enumerate}}
\def\een{\end{enumerate}}

\def\Maxlp{\raisebox{-1.5 ex}{\rlap{\tiny $\;\;p_i^2\le c\L^2$}}
\raisebox{0ex} {$\; \mbox{Max}\;\;\;\,$}}
\def\nome#1{{\label{#1}}}
\def\PD {\partial}

\begin{document}
\begin{titlepage}
\renewcommand{\thefootnote}{\fnsymbol{footnote}}
\begin{flushright}
     UPRF 93-388 \\
     October 1993
\end{flushright}
\par \vskip 10mm
\begin{center}
{\Large \bf
Renormalization group flow for $SU(2)$ Yang-Mills\\
theory and gauge invariance\footnote{
Research supported in part by MURST, Italy}}
\end{center}
\par \vskip 2mm
\begin{center}
        {\bf M.\ Bonini, M.\ D'Attanasio and G.\ Marchesini} \\
        Dipartimento di Fisica, Universit\`a di Parma and\\
        INFN, Gruppo Collegato di Parma, Italy
        \end{center}
\par \vskip 2mm
\begin{center} {\large \bf Abstract} \end{center}
\begin{quote}
We study the formulation of the Wilson renormalization group (RG)
method for a non-Abelian gauge theory.
We analyze the simple case of $SU(2)$ and show that
the local gauge symmetry can be
implemented by suitable boundary conditions for the RG flow.
Namely we require that the relevant couplings present in the
physical effective action, \ie the coefficients of the field monomials
with dimension not larger than four, are fixed to satisfy the
Slavnov-Taylor identities. The full action obtained from the RG
equation should then satisfy the same identities.
This procedure is similar to the one we used in QED. In this way we
avoid the cospicuous fine tuning problem which arises if one gives
instead the couplings of the bare Lagrangian.
To show the practical character of this formulation we deduce
the perturbative expansion for the vertex functions in terms
of the physical coupling $g$ at the subtraction point $\mu$
and perform one loop calculations.
In particular we analyze to this order some ST identities and compute
the nine bare couplings.
We give a schematic proof of perturbative renormalizability.
\end{quote}
\end{titlepage}

\section{Introduction}
An elegant treatment of local gauge theories is provided by
dimensional regularization \cite{eps} which preserves BRS invariance.
However, this method is necessarily bounded to the perturbative
regime. Moreover, in the case of chiral gauge theories, even this
regularization conflicts \cite{chi} with the local symmetry. The
reason for this is that the usual definition of $\g_5$
is not consistent in complex dimension. Therefore it is useful to study
and exploit alternative formulations which work in four dimensions and
are in principle nonperturbative.

The exact renormalization group (RG) formulation \cite{W} works in the
physical space-time and provides a deep physical meaning of the
ultraviolet divergences. The original motivation for this method was
the study of the flow of Lagrangians at various scales. Once the
action $S_{\L_0}$ at the ultraviolet (UV) scale $\L_0$ is given, one
computes the Wilsonian action $S_\L$ at a lower scale $\L$, by integrating the
fields with frequencies $p$ between $\L$ and $\L_0$ ($\L^2 <p^2
<\L_0^2$). A very simple way of constructing the RG equations has been
written few years ago by Polchinski \cite{P}. See also
\cite{G}-\cite{oth}.
This method not only allows the study of the evolution in $\L$ of
$S_\L$, but provides a practical way of computing Green's functions.
For instance the iterative solution of RG equations gives the usual
loop expansion.

The RG formulation was not considered suitable for local gauge
theories since the presence of a momentum scale $\L$ implies that the BRS
invariance is completely lost. As a consequence the bare couplings of
the UV action $S_{\L_0}$ are not constrained by BRS invariance.
Obviously the various bare couplings in $S_{\L_0}$ are anyway
related by the fact that the physical Green's functions must
satisfy Slavnov-Taylor (ST) identities. Even assuming that this is
possible, the bare couplings would be constrained by a tremendous
fine tuning process \cite{romani}.
Recently Becchi \cite{B}, by studying the ``quantum action principle''
for the UV action in $SU(2)$ Yang-Mills (YM) case was able to extend
the BRS transformations and obtain a functional fine tuning equation
for the bare couplings. See also \cite{Riv}.

In this paper we present an attempt of giving a RG formulation for
gauge theories which avoids the fine tuning problem.
In a previous paper \cite{QED} we discussed the Abelian case while
here we analyze the non-Abelian case of $SU(2)$ Yang-Mills theory.
The method is quite general, uses some well known procedures but organized
in a rather unusual way. Therefore it is useful
to summarize the main steps.
\newline{\it 1) Definition of the cutoff effective action.}
Consider the physical effective action $\G[\Phi]$ where $\Phi(x)$
denotes all the fields of the theory. For QED $\Phi=(A_\mu,\bar
\psi,\psi)$, \ie the photon and electron fields; for SU(2) YM case
$\Phi=(A_\mu^a,c^a,\bc^a,u_\mu^a,v^a)$, \ie the gauge, ghost,
antighost fields and the sources for the BRS transformations of
$A_\mu^a$ and $c_a$ respectively. $\G[\Phi]$ must satisfy Ward or ST
identities. We consider now the ``cutoff effective action'' $\G[\Phi;\L]$,
obtained by putting an infrared (IR) cutoff $\L$ and an UV cutoff
$\L_0$ for all propagators in the vertices, \ie we set to zero each
propagator if its frequency is lower than $\L$ and larger than $\L_0$.
In the \UV limit $\G[\Phi;\L=0]$ is just the physical effective action,
while for $\L\ne 0$ it does not satisfy Ward or ST
identities. The functional $\G[\Phi;\L]$ is related to the Wilsonian
action $S_\L$ and we introduced it only because of technical
reasons. Actually $\G[\Phi;\L=\L_0]$ is the UV action $S_{\L_0}$.
\newline{\it 2) RG equation for} $\G[\Phi;\L]$. The RG equation for
$S_\L$ can be formulated directly for $\G[\Phi;\L]$ and, as we shall
discuss, takes the form (see \cite{P})
\beq\nome{1}
\L\partial_\L \G[\Phi;\L]=I[\Phi;\L]
\eeq
where the functional $I[\Phi;\L]$ is given (non linearly) in terms of
$\G[\Phi;\L]$. Once the boundary conditions are given, the RG equation
\re{1} allows one to obtain $\G[\Phi;\L]$ for any $\L$ and in particular,
for $\L=0$, the physical effective action.
As one expects, the boundary conditions for the various vertices
depend on dimensional counting.
\newline{\it 3) Relevant and irrelevant parts of} $\G[\Phi;\L]$.
One distinguishes irrelevant vertices, which have negative mass
dimension, and relevant couplings with non-negative dimension. One can
therefore write
\beq\nome{2}
\G[\Phi;\L]=\G_{rel}[\Phi;\s_i(\L)] + \G_{irr}[\Phi;\L]\,,
\eeq
where $\G_{irr}[\Phi;\L]$ involves only the irrelevant vertices and
$\s_i(\L)$ are the relevant couplings: seven for QED and nine for the
$SU(2)$ case. Therefore $\G_{rel}$ is given by a polynomial in the
fields. The relevant couplings are defined as the values of some
vertex functions or their derivatives at a given normalization point.
We are ready now to discuss the boundary conditions for the evolution
equation in \re{1}.
\newline{\it 4) Boundary conditions for $\G_{irr}[\Phi;\L]$}.
For $\L=\L_0\to\infty$ one assumes
\beq\nome{3}
\G[\Phi;\L_0]=\G_{rel}[\Phi;\s_i(\L_0)]\,,
\eeq
\ie the irrelevant vertices vanish in the UV region. Notice that this
hypothesis is essential to have perturbative renormalizability.
The parameters $\s_i(\L_0)$ are the bare parameters of the UV action.
In the usual procedure one gives these bare parameters and,
by using \re{1}, evaluates $\G[\Phi;\L=0]$.
The requirement that this resulting effective action  satisfies the ST
identities gives rise to the fine tuning problem, which we want to
avoid.
\newline{\it 5) Boundary conditions for  $\G_{rel}[\Phi;\s_i(\L)]$}.
Instead of fixing the boundary conditions for the
relevant parameters $\s_i(\L)$ at $\L=\L_0\to\infty$, we set them at
the physical point $\L=0$. The values $\s_i(0)$ are then the physical
masses, wave function costants and couplings. Moreover, for a gauge
theory we must fix these couplings in such a way that
$\G_{rel}[\Phi;\s_i(0)]$ satisfies Ward or ST identities.
This is then the place where we implement the symmetry of the theory.
For the Abelian case, studied in the previuos paper, we showed that
one can simply assume $\G_{rel}[\Phi;\s_i(0)]=S_{cl}[\Phi]$, where
$S_{cl}[\Phi]$ is the QED classical action including the gauge fixing
term. Then $\G_{rel}[\Phi;\s_i(0)]$ satisfies Ward identities.
For the non-Abelian theory, as we shall illustrate, the situation is more
complex.
In this case one has two features which are new
with respect to QED.
The first difference is that in the YM case one can construct relevant
monomials, such as for instance $(A_\mu^a A^a_\mu)^2$,
which are not present in the classical Lagrangian.
The second important difference between QED and YM theories is that
Ward identities are linear while ST identities are not linear.
In QED, Ward identities do not couple relevant and irrelevant
couplings, so that the relevant part of the effective action
satisfies Ward identities by itself.
In the YM case instead the ST identities couple the relevant and
irrelevant part of the effective action. Therefore the requirement
that $\G_{rel}[\Phi;\s_i(0)]$ satisfies ST identities implies thas
some of the nine relevant couplings are fixed in terms of irrelevant
vertices. This will be discussed in detail in sect.~3.
\newline{\it 6) Analysis of the ST identities}.
These boundary conditions and the RG equation \re{1} completely
determine $\G[\Phi;\L]$, at least perturbatively. The fundamental
point is whether the physical effective action $\G[\Phi;\L=0]$ does
satisfy Ward or ST identities. Recall that these identities have been
implemented only for the relevant part of $\G[\Phi;\L=0]$.
In the Abelian case we have been able to show \cite{QED} that the
renormalized Green's functions indeed satisfy Ward identities to all
order in perturbation theory.
The method was based on the properties of RG flow and the use
of the above boundary conditions. However the proof still required the
analysis of graphs. The same method could be extended to
the non-Abelian case but it would be quite more complicated.
In this paper we limit ourself
to the analysis of some ST identities to one loop. In this way we
illustrate the key points needed to obtain these identities.
A general proof to all loops should use more synthetic techniques.

The paper is organized as follows.
In sect.~2
we define the cutoff effective action for the $SU(2)$ YM theory
for which we write the RG equation.
In sect.~3
we define the relevant couplings and deduce in detail their values at
$\L=0$ to satisfy ST identities.
In sect.~4
we convert the differential functional equation into an integral
equation which embodies the boundary conditions.
In sect.~5
we show that the iterative solution
gives the loop expansion for the vertex functions. This perturbative
expansion involves only the physical parameter $g$ defined at the
subtraction point $\mu$.
To one-loop order we explicitly derive:
the various vertex functions,
the bare parameters in the UV Lagrangian, \ie the relevant couplings
$\s_i(\L)$ at $\L=\L_0$, and
the usual one-loop beta function.
We check the ST identities at one loop
for the vector propagator and for the ghost-vector vertex in the
physical limit ($\L=0$ and \UV).
Finally we give a schematic proof of perturbative renormalizability.
In sect.~6 we show that the cutoff effective action depends, as the
classical Lagrangian, on the combination $w^a_\mu=u^a_\mu/g+\PD_\mu
\bc^a$. This fact simplify greatly the analysis of relevant field
monomials.
Sect.~8 contains some conclusions.

\section{Renormalization group flow and effective action}
In this section we introduce the cutoff effective action
and deduce the RG group flow equations. We follow the same
steps \cite{BDM,QED} as in the scalar and QED theory.
This allows us to set up the necessary notations.

In the $SU(2)$ gauge theory the classical Lagrangian in the
Feynman gauge is
\beq\nome{Scl}
S_{cl}[A_\mu,c,\bc]
=\int d^4x
\left\{
-\frac 1 4 \left( F_{\mu\nu}  \cdot F_{\mu\nu}\right)
-\half  \left(\PD_\mu A_\mu \right)^2
- \bc \cdot \PD_\mu D_\mu c
\right\}\,,
\eeq
where
$$
F^a_{\mu\nu}(x)=\PD_\mu A^a_\nu -\PD_\nu A^a_\mu
+ g \left( A_\mu \wedge A_\nu\right)^a \,,
\;\;\;\;\;\;
\left(A_\mu \wedge A_\nu\right)^a=
\epsilon^{abc} A^b_\mu A^c_\nu \,,
$$
$$
F_{\mu\nu} \cdot F_{\mu\nu} =F_{\mu\nu} ^a F_{\mu\nu} ^a\,,
\;\;\;\;\;\;
D_\mu c =\PD_\mu c + gA_\mu \wedge c\,.
$$
This action is invariant under the BRS transformations \cite{BRS}
$$
\delta A_\mu= \frac 1 g \eta D_\mu c
\,,
\;\;\;\;\;\;\;
\delta c= -\eta \half c \wedge c
\,,
\;\;\;\;\;\;\;
\delta \bc= - \frac 1 g \eta \PD_\mu A_\mu
\,,
$$
with $\eta$ a Grassmann parameter.
Introducing the sources $u_\mu$ and $v$ associated to the variations
of $A_\mu$ and $c$ one has the BRS action
\beq\nome{Stot}
S_{BRS}[A_\mu,c,\bc,u_\mu,v]
=
S_{cl} +
\int d^4x
\left\{ \frac 1 g u_\mu  \cdot D_\mu c -\half v \cdot c \wedge c
\right\}=S_2+S_{int}\,,
\eeq
with
\beq\nome{S2}
S_2=\int d^4 x \left\{
\half A_\mu\cdot\partial^2  A_\mu +
w_\mu \cdot \partial_\mu c
\,- \half
v\cdot c\wedge c \right\}
\eeq
and
\begin{eqnarray*}
&&
S_{int}= \int d^4x \left\{ig\;
(\partial_\nu A_\mu) \cdot A_\nu \wedge A_\mu
-g \; w_\mu \cdot c \wedge A_\mu
-\frac {g^2}{4}\;
(A_\mu\wedge A_\nu)\cdot (A_\mu\wedge A_\nu) \right\}
\,.
\end{eqnarray*}
These expressions are functions of the combination
$w_\mu=\frac 1 g u_\mu+\PD_\mu \bc$.
One expects that also the effective action depends on this
combination of fields.
In sect.~6 we will show that this property holds also for
the interacting part of the cutoff effective action.

\subsection{Cutoff effective action}
We now define for this theory the cutoff effective action discussed
in the introduction.
In order to compute the vertices one
needs a regularization procedure of the ultraviolet divergences.
We regularize these divergences by assuming that in the path integral
one integrates only the fields with frequencies smaller than a given
UV cutoff $\L_0$. This procedure is equivalent to assume that the free
propagators vanish for $p^2 > \L_0^2$.
The physical theory is obtained by considering the limit \UV.
In order to study the Wilson renormalization group flow \cite{W,P},
one introduces in the free propagators also an infrared cutoff $\L$.
The quadratic part of the action \re{S2} becomes, in momentum space,
\begin{eqnarray*}
S_2^{\L,\L_0}= & \int_p  \;
\left\{
-\half A_\mu(-p)\cdot A_\mu(p) \, p^2  \;+\; p^2 \bc(-p)\cdot c(p)
\right\}
\,K^{-1}_{\L\L_0}(p)
\\ &
-\frac {i}{g} \int_p  p_\mu u_\mu(-p)\cdot c(p)
\,- \frac 1 2
\int_p\int_q v(p)\cdot c(q)\wedge c(-p-q)
\,,
\end{eqnarray*}
where $K_{\L\L_0}(p)=1$ in the region
$\L^2 \, \ltap \,p^2\, \ltap \, \L_0^2$ and rapidly vanishing outside
and
$$
\int_p \equiv \int \frac {d^4p}{(2\pi)^4}\,.
$$
The introduction of a cutoff in the propagators breaks the gauge
invariance properties of the action.
Therefore at the UV scale the interaction
$S_{int}^{\L_0}$ must contain all
the nine monomials which are $SU(2)$ singlets and Lorentz scalar
with dimension not higher than four. The first seven monomials
\beq\nome{7s}
\eqalign{
&
A_\mu \cdot A_\mu \,,
\;\;\;\;\;\;\;
(\partial_\nu A_\mu) \cdot (\partial_\nu A_\mu) \,,
\;\;\;\;\;\;\;
(\partial_\mu A_\mu) \cdot (\partial_\nu A_\nu) \,,
\;\;\;\;\;\;\;
w_\mu \cdot \partial_\mu c \,,
\cr &
(A_\mu\wedge A_\nu) \cdot (\partial_\mu A_\nu) \,,
\;\;\;\;\;\;\;
w_\mu \cdot c \wedge A_\mu \,,
\;\;\;\;\;\;\;
(A_\mu \wedge A_\nu)\cdot (A_\mu \wedge A_\nu)
}
\eeq
are present in the classical action \re{Scl}.
The remaining two
\beq\nome{2r}
2(A_\mu \cdot A_\nu)\, (A_\mu \cdot A_\nu)
\,+\,
(A_\mu \cdot A_\mu)\, (A_\nu \cdot A_\nu) \,,
\;\;\;\;\;\;
v \cdot c \wedge c
\eeq
are not present in \re{Scl} and are generated by the interaction.

In general we consider two functionals $\Pi_{1;rel}$ and
$\Pi_{2;rel}$, containing the two groups of monomials in
\re{7s} and \re{2r} respectively.
The first one depends on the seven couplings $\s_i$
\beq\nome{Pis}
\eqalign{
& \Pi_{1;rel}[\Phi;\s_i]
=\int d^4x \biggl\{
\frac 1 2 A_\mu \cdot
\left[g_{\mu\nu}(\s_{m_A} -\s_\alpha \partial^2)
- \s_A (g_{\mu\nu} \partial^2-\partial_\mu \partial_\nu)\right] A_\nu+
i\s_{wc}\; w_\mu \cdot \partial_\mu c
\cr \;\;& +
i\s_{3A}\;(\partial_\nu A_\mu) \cdot A_\nu \wedge A_\mu+
\frac{\s_{4A}}{4}\;(A_\mu\wedge A_\nu)\cdot(A_\mu\wedge A_\nu)+
\s_{wcA} \;w_\mu\cdot c\wedge A_\mu \biggr\}\,.
}
\eeq
The second functional depends on two additional couplings.
Since they will play a particular r\^ole, we denote them with the
different letter $\r$. The functional $\Pi_{2;rel}$ is then
\beq\nome{Pir}
\Pi_{2;rel}[\Phi;\r_i]
=\int d^4x \left\{
\frac{\r_{vcc}}{2}\;
v\cdot c \wedge c +
\frac{\r_{4A}}{8}\;
[2\,(A_\mu\cdot A_\nu)\,(A_\mu\cdot A_\nu)\, +
(A_\mu\cdot A_\mu)\,(A_\nu\cdot A_\nu)]
\right\}
\eeq
and contains the contributions from field monomials
in \re{2r}, which do not appear in the classical action.
By $\Phi$ we indicate the classical fields and sources
$A_\mu$, $c$, $w_\mu$ and $v$.
In conclusion the interacting part of the UV action can be written as
\beq\nome{intac}
S_{int}^{\L_0}=\Pi_{1;rel}[\Phi;\s_i^B]
+\Pi_{2;rel}[\Phi;\r_i^B]
\,,
\eeq
where
$\s_i^B$ and $\r_i^B$ are the bare couplings.
The mass parameter $\s^B_{m_A}$ has positive dimension, while the other
parameters are dimensionless. $\s^B_A$ and $\s^B_{wc}$ are related to the
vector and ghost wave function renormalization,
$\s^B_\alpha$ to the gauge fixing parameter renormalization,
$\s^B_{wcA}$,
$\s^B_{3A}$ and
$\s^B_{4A}$
are related to the bare interaction couplings.
The additional coupling $\r^B_{4A}$ corresponds to the interaction among
four vectors with a group structure not present in the classical
action \re{Scl}.
Similarly $\r^B_{vcc}$ is an interaction term not present in $S_{cl}$.

The generating functional of the Green's functions is
$$
Z[j,u,v;\L]=\exp i W[j,u,v;\L] =\,
\int D \left[ \phi_A \right]\;
\exp i\left\{ S_2^{\L,\L_0}+ S_{int}^{\L_0} +
(j,\phi) \right\} \,,
$$
where we introduced the compact notation for the fields and
corresponding sources
$$
\phi_A=(A^a_\mu,\,c^a,\,\bc^a )\, ,
\;\;\;\;\;\;\;\;
j_A=(j^a_\mu,\, \bar \chi^a ,\, - \chi^a )\,,
$$
$$
(j,\phi) \equiv \int_p j_\mu(-p) \cdot A_\mu(p)
+ \bar \chi(-p) \cdot c(p)
+ \bc(-p) \cdot \chi(p)
\,.
$$
The free cutoff propagators are described by the matrix $D_{AB}$
defined by
$$
\int_p\half \phi_A(-p) D^{-1}_{AB}(p;\L)\phi_B(p)
\equiv
\int_p  \;
\left\{
-\half A_\mu(-p)\cdot A_\mu(p) \, p^2  \;+\; p^2 \bc(-p)\cdot c(p)
\right\}
\,K^{-1}_{\L\L_0}(p)\,.
$$
In $Z[j,u,v;\L]$  and $D_{AB}(p;\L)$ we have
explicitly written only the cutoff
$\L$ since we will consider in any case the limit \UV.
The physical functional $Z[j,u,v]$ is obtained by taking
the limits \IR and \UV.

Finally, the cutoff effective action is defined by taking the Legendre
transform
$$
\G[\phi,u,v;\L]=W[j,u,v;\L] - W[0,0,0;\L] - (j,\phi) \, ,
\;\;\;\;\,\;\;
\phi_A(p) = (2\pi)^4  \frac {\delta W[j,u,v;\L] }{\delta j_A(-p)} \, .
$$

\subsection{Exact renormalization group equation}
Usually the RG equations are obtained by requiring that the
physical quantities are independent of $\L$. In the present
formulation the same flow equation can be simply obtained by observing
that all the $\L$ dependence of $Z[j,u,v;\L]$ is contained in the cutoffs
in the propagators. Thus one easily derives \cite{P} the equation
$$
\dL Z[j,u,v;\L]=-i\frac{ (2\pi)^8}{2}\; \int_q\;\dL{D^{-1}_{BA}(q;\L)}
\frac {\delta^2 Z[j,u,v;\L] }{\delta j_B(q) \delta j_A(-q) } \, .
$$
In the same way one finds the corresponding equation for the cutoff
effective action. The equation is schematically given in fig.~1 and
is obtained by observing that the operator $\L\partial_\L$ acts on each
internal propagator. The first term involves only one vertex;
the second term involves two vertices in such a way to reproduce,
upon $q$-integration, a one-particle irreducible contribution. The
other terms will involve any possible number of vertices.
Of course one can deduce formally this equation. By following the same
steps of refs. \cite{BDM,QED} we obtain
\beq\nome{eveq}
\dL \Pi[\Phi;\L]
=
- \frac i 2 \int_q M_{BA}(q;\L)\; \bG_{AB}[-q,q;\Phi;\L]
\equiv I[\Phi;\L]\,,
\eeq
where
\beq\nome{gamma}
\G[\phi,u,v;\L]=S_2^{\L,\L_0}+\Pi[\Phi;\L]\,,
\eeq
$$
M_{BA}=\Delta_{BC}(q;\L) \dL{ D_{CD}^{-1}(q;\L)} \Delta_{DA}(q;\L)
$$
and $\Delta_{AB}(q;\L)$ is the full propagator.
The auxiliary functional $ \bG_{AB}[q,q';\Phi]\,$
is the inverse of
${\delta^2\G[\phi,u,v]}/ {\delta\phi_A(q)\delta\phi_B(q')}$.
To obtain this we isolate the contribution of the two point function
$$
(2\pi)^8\frac {\delta^2\G[\phi,u,v]} {\delta\phi_A(q)\delta\phi_B(q')}
=
(2\pi)^4\delta^4(q+q') \, \Delta^{-1}_{BA}(q;\L)+
\G_{BA}^{int}[q',q;\Phi]\,.
$$
The auxiliary functional $ \bG_{AB}[q,q';\Phi]\,$ is then given by
the integral equation
\beq \nome{bG}
\bG_{AB}[q,q';\Phi] = (-)^{\delta_B}\,\G_{AB}^{int}[q,q';\Phi]\,-
\int_{q''} \bG_{CB}[-q'',q';\Phi]
\Delta_{DC}(q'';\L)
\G_{AD}^{int}[q,q'';\Phi]
  \, ,
\eeq
where $\delta_B$ is the ghost number and
the indices $A$ and $B$ run over the indices of the fields
$A^a_\mu$, $\bc^a$ and $c^a$.
In terms of the proper vertices
of $\Pi[\Phi;\L]$ the evolution equations are
$$
\dL \Pi_{C_1\cdots C_n}(p_1, \cdots p_{n};\L)
=  -\frac i 2 \int_q M_{BA}(q;\L)\;
\bG_{AB,C_1\cdots C_n}(-q,q;p_1, \cdots p_{n};\L) \,,
$$
where $\bG_{AB,C_1\cdots C_n}(q,q';p_1, \cdots p_{n};\L)$
are the vertices of the auxiliary functional $\bG[\Phi;\L]$
obtained by differentiating with respect to the fields
$\Phi$.
The vertices of this auxiliary functional are obtained
in terms of the proper vertices by expanding \re{bG} and one finds
\beq
\eqalign{
&
\bG_{AB, C_1\cdots C_n} (q,q';p_1,\cdots p_{n};\L)=
 \G_{A B C_1\cdots C_n} (q,q',p_1,\cdots p_{n};\L)
\cr&
-{\sum}'
\G_{A C_{i_1}\cdots C_{i_{k}}C'} (q,p_{i_1},\cdots,p_{i_{k}},Q;\L)
\;\Delta_{C'C''}(Q;\L)
\bG_{C''B,C_{i_{k+1}}\cdots C_{i_{n}}}
(-Q,q';p_{i_{k+1}},\cdots p_{i_{n}};\L) \,,
}
\eeq
where $Q=q+p_{i_1}+\cdots p_{i_{k}}$,
and $\sum'$ is the sum over the combinations of photon and
ghost indices $(i_1 \cdots i_{n})$ taking properly into account the
symmetrization and anti-symmetrization.
Notice that this construction involves vertices with $\bc$ external fields,
which can be obtained from the relation
$\G_{\bc\cdots}(p,\cdots)=-ip_\mu\G_{w_\mu \cdots}(p,\cdots)$.

\section{Relevant parameters and physical conditions}
In order to define the boundary conditions for the evolution
equation of the cutoff effective action
we have to distinguish the relevant and irrelevant vertices.
The interaction part $\Pi[\Phi;\L]$ in \re{gamma} can be written as
$$
\Pi[\Phi;\L]=
 \Pi_{1;rel}[\Phi;\s_i(\L)]
+\Pi_{2;rel}[\Phi;\r_i(\L)]
+\G_{irrel}[\Phi;\L] \,,
$$
where the functional $\G_{irrel}$ contains all irrelevant vertices.
The relevant part of $\Pi[\Phi;\L]$ involves only field monomials with
dimension not larger than four. Therefore the structure of $\Pi_{1,2;rel}$
is the same as the one defined in \re{Pis} and \re{Pir} for the UV
action. Now the couplings $\s_i$ and $\r_i$ depend on $\L$. These nine
relevant couplings are defined as the value of some vertex functions
or their derivatives at a given normalization point. In order to
split the functional $\Pi[\Phi;\L]$ in its relevant and irrelevant
parts, it is then necessary to specify the procedure of extracting the
relevant parameters. This is done in full detail in the next
subsection.

\subsection{Relevant couplings and subtraction conditions}
In this subsection we define the couplings $\s_i(\L)$ and $\r_i(\L)$
as the values at the subtraction points of the vertices with
non-negative dimension. We work in the momentum space.
The subtraction point for the two point functions is assumed
at $p^2=\mu^2$,
while for the $N$ point vertex functions is assumed at the symmetric point
($NSP$) defined by
$$
\bar p_i \bar p_j=\frac{\mu^2}{N-1}(N\delta_{ij}-1)\,,
\;\;\;\;\;\;\;\;\;N=3,4,...\,.
$$
The relevant couplings are involved in the following six
$SU(2)$ singlets vertices with
$$
\dim \le 4\,,
$$
where $n_i$ is the number of fields of the type $i$.

\noindent
1) The vector propagator is the coefficient of
$\half A_\mu(-p)\cdot A_\nu(p)$ and has the form
\beq\nome{propp}
\G_{\mu\nu}(p;\L)=
-g_{\mu\nu}\, p^2 \,K_{\L\L_0}^{-1}(p) +
g_{\mu\nu}\Pi_L(p;\L) + t_{\mu\nu}(p)\,\Pi_T(p;\L)\,,
\eeq
with
$$
t_{\mu\nu}(p)\equiv g_{\mu\nu}\;-\;\frac {p_\mu p_\nu}{p^2} \,.
$$
The three relevant couplings are defined by
\beq\nome{sl}
\Pi_L(p;\L)=\s_{m_A}  (\L)+p^2 \s_\alpha(\L) + \Sigma_L(p;\L)\,,
\;\;\;\;\;
\Sigma_{L}(0;\L)=0\,,
\;\;\;\;\;
\frac{\partial \Sigma_{L}(p;\L)}{\partial p^2}|_{p^2=\mu^2}=0\,,
\eeq
and
\beq\nome{st}
\Pi_T(p;\L)=p^2 \s_A(\L) + \Sigma_T(p;\L)\,,
\;\;\;\;
\Sigma_{T}(0;\L)=0\,,
\;\;\;\;
\frac{\partial \Sigma_{T}(p;\L)}{\partial p^2}|_{p^2=\mu^2}=0\,.
\eeq
{}From these conditions we can factorize in the vertices $\Sigma_{L,T}$
a dimensional function of $p$. Thus $\Sigma_{L,T}$ are ``irrelevant''
and contribute to $\G_{irrel}[\L]$.

\par\noindent
2) The contribution to $\G[\L]$ from three vectors can be written as
$$
A_\mu(p)\cdot A_\nu(q)\wedge A_\r(k)
\; \G_{\mu\nu\r}^{(3A)}(p,q,k;\L)\,,
$$
\beq\nome{3A}
\G_{\mu\nu\r}^{(3A)}(p,q,k;\L)=
p_\nu g_{\mu\r} \G^{(3A)}(p,q,k;\L)
+
\tilde \G_{\mu\nu\r}^{(3A)}(p,q,k;\L)
\,.
\eeq
In the last term all three Lorentz indices are carried by external
momenta. Thus, after factorizing these momenta, one remains with a
function of dimension $-2$. Then the vertex
$\tilde \G_{\mu\nu\r}^{(3A)}(p,q,k;\L)$ is irrelevant and
contributes to $\G_{irrel}$.
The relevant coupling $\s_{3A}(\L)$ is defined by
$$
\G^{(3A)}(p,q,k;\L)= \s_{3A}(\L)+\S^{(3A)}(p,q,k;\L)\,,
\;\;\;\;\;\;\;
\S^{(3A)}(p,q,k;\L)|_{3SP}=0 \,,
$$
so that $\S^{(3A)}(p,q,k;\L)$ is irrelevant.

\par\noindent
3) The contribution to $\G[\L]$ from four vectors is given by two
different $SU(2)$ scalars. The first one involves the field combination
which appears in the classical action:
$$
\frac{1}{4} [A_\mu(p)\wedge A_\nu(q)]
\cdot
[A_\r(k)\wedge A_\s(h)]
\, \G_{1;\mu\nu\r\s}^{(4A)}(p,q,k,h;\L)\,,
$$
\beq\nome{4A1}
\G_{1;\mu\nu\r\s}^{(4A)}(p,q,k,h;\L)=
g_{\mu\r} g_{\nu\s}\, \G_1^{(4A)}(p,q,k,h;\L)
+
\tilde \G_{1;\mu\nu\r\s}^{(4A)}(p,q,k,h;\L)
\,.
\eeq
The other singlet structure is given by
$$
\frac 1 8 [A_\mu(p)\cdot A_\nu(q)]\; [A_\r(k)\cdot A_\s(h)]
\, \G_{2;\mu\nu\r\s}^{(4A)}(p,q,k,h;\L)\,,
$$
\beq\nome{4A2}
\G_{2;\mu\nu\r\s}^{(4A)}(p,q,k,h;\L)
=
(2g_{\mu\r} g_{\nu\s}\,+ g_{\mu\nu} g_{\r\s}\,)\;
\G_2^{(4A)}(p,q,k,h;\L)
+
\tilde \G_{2;\mu\nu\r\s}^{(4A)}(p,q,k,h;\L)
\,.
\eeq
For the two vertices $\tilde \G_{1,2;\mu\nu\r\s}^{(4A)}$
at least two Lorentz indices are carried by external
momenta. Thus these vertices are irrelevant.
The two relevant couplings $\s_{4A}(\L)$ and $\r_{4A}(\L)$ are defined by
$$
\G_1^{(4A)}(p,q,k,h;\L)= \s_{4A}(\L)+\S_1^{(4A)}(p,q,k,h;\L)\,,
\;\;\;\;\;
\S_1^{(4A)}(p,q,k,h;\L)|_{4SP}=0\,,
$$
$$
\G_2^{(4A)}(p,q,k,h;\L)= \r_{4A}(\L)+\S_2^{(4A)}(p,q,k,h;\L)\,,
\;\;\;\;
\S_2^{(4A)}(p,q,k,h;\L)|_{4SP}=0\,.
$$

\par\noindent
4) The interaction part of the ghost propagator and $u$-$c$ vertex
is given by the same function.
The ghost propagator is the coefficient of $\bc(-p)\cdot  c(p)$
\beq\nome{bcc}
\G^{(\bc c)}(p;\L)=
p^2K_{\L\L_0}^{-1}(p) + i p^2 \Pi^{(wc)}(p;\L)
\,.
\eeq
The coefficient of $u_\mu(-p)\cdot c(p)$ has the form
\beq\nome{uc}
\G_\mu^{(uc)}(p;\L)=
-\frac{ip_\mu}{g} +\frac{p_\mu}{g} \Pi^{(wc)}(p;\L) \,.
\eeq
The function $\Pi^{(wc)}$ contains the relevant coupling $\s_{wc}(\L)$
\beq\nome{wc}
\Pi^{(wc)}(p;\L)=
\s_{wc}(\L)+\S^{(wc)}(p;\L) \,,
\;\;\;\;\;\;
\partial_{p^2} \left( p^2\S^{(wc)}(p;\L) \right)|_{p^2=\mu^2}
=0\,.
\eeq

\par\noindent
5) The contribution to $\G[\L]$ from $w$-$c$-$A$ is given by the
$SU(2)$ scalar
$$
w_\mu(p)\cdot c(q)\wedge A_\nu(k)\;
\, \G_{\mu\nu}^{(wcA)}(p,q,k;\L)\,,
$$
\beq\nome{wca?}
\G_{\mu\nu}^{(wcA)}(p,q,k;\L)=
g_{\mu\nu}\,\G^{(wcA)}(p,q,k;\L)
+
\tilde \G_{\mu\nu}^{(wcA)}(p,q,k;\L)\,.
\eeq
The two Lorentz indices in $\tilde \G_{\mu\nu}^{(wcA)}(p,q,k;\L)$
are carried by external momenta. Thus this vertex is irrelevant.
The coupling $\s_{wcA}(\L)$ is defined by
\beq\nome{wcA}
\G^{(wcA)}(p,q,k;\L) = \s_{wcA}(\L) +
\S^{(wcA)}(p,q,k;\L)\,,
\;\;\;\;\;
\S^{(wcA)}(p,q,k;\L)|_{3SP}=0 \,.
\eeq
Due to the fact that $\G$ depends on $\bc$ and $u$ only through the
combination $w_\mu$ one has that the coefficient of $\bc$-$c$-$A$ is
\beq
\nome{bccA}
\G_{\nu}^{(\bc cA)}(p,q,k;\L)=-ip_\mu\,
\, \G_{\mu\nu}^{(wcA)}(p,q,k;\L)\,.
\eeq

\par\noindent
6) The only vertex involving the source $v$
which contains a relevant coupling is the coefficient of
$\half v(p)\cdot c(q) \wedge c(k)$. The coupling $\r_{vcc}$ is given by
$$
\G^{(vcc)}(p,q,k;\L)=-1+\r_{vcc}(\L)+ \S^{(vcc)}(p,q,k;\L)\,,
\;\;\;\;
\S^{(vcc)}(p,q,k;\L)|_{3SP}=0\,.
$$
All the remaining vertices are irrelevant, since they are coefficients of
monomials with dimension higher than four.

\subsection{Boundary conditions}
As discussed in the introduction for the irrelevant vertices
we assume the following boundary condition
\beq\nome{bcirr}
\G_{irr}[\Phi,\L_0] =0 \,,
\eeq
since, due to dimensional reasons, they vanish at the UV scale.
For $\L=\L_0$ the cutoff effective action becomes then local and
corresponds to the bare action \re{intac}, where
the nine bare parameters $\s_i^B$ and $\r_i^B$ are given by
$\s_i(\L_0)$ and $\r_i(\L_0)$. However, in our procedure these
couplings are fixed at $\L=0$ and their value at $\L=\L_0$ can be
perturbatively computed.

For the relevant part, we first give the boundary conditions for the
terms which appear also in the classical action.
For them we impose
$$
\Pi_{1;rel}[\Phi;\s_i(\L=0)] =S_{int}\,,
$$
which gives
\beq\nome{bcs}
\eqalign{
&
\s_{m_A}(0)=\s_{\alpha}(0)=\s_{A}(0)=\s_{wc}(0)=0\,,
\cr&
\s_{3A}(0)=ig
\,,\;\;\;
\s_{4A}(0)=-g^2
\,,\;\;\;
\s_{wcA}(0)=-g\,.
}
\eeq
The remaining two couplings $\r_{4A}(0)$ and $\r_{vcc}(0)$ are fixed by
the ST identities and given in terms of some irrelevant vertices as
follows.

The physical effective action, namely $\G[\Phi;\L=0]$, must satisfy the
ST identities (in the rest of this subsection it will be understood
that $\L=0$ and \UV).
Since, as discussed, the effective action depends on $\bc$ and $u_\mu$
fields only through the combination $w_\mu$, the Slavnov-Taylor identities
are
\beq\nome{ST}
\int_s \;
\left\{
\frac{\delta\G'}{\delta u_\mu^a(s)}
\frac{\delta\G'}{\delta A_\mu^a(-s)}
+
\frac{\delta\G'}{\delta v^a(s)}
\frac{\delta\G'}{\delta c^a(-s)}
\right\}
=0 \,,
\eeq
where  $\G'$ is obtained from
the effective action $\G$ by subtracting the gauge fixing contribution,
\ie the second term in \re{Scl}.

These relations, due to their non-linearity, couple the relevant and
irrelevant vertices.
Since the total action $S_{BRS}$ (without the gauge fixing term)
satisfies these identities,
only the relevant couplings $\r_{4A}(0)$ and $\r_{vcc}(0)$ are
coupled to irrelevant vertices.
We now determine the relations giving the couplings
$\r_{4A}(0)$ and $\r_{vcc}(0)$.

To find the relation for $\r_{vcc}(0)$ we use the simplest identity
involving the $v$-$c$-$c$ vertex. This is obtained by
differentiating \re{ST} with respect to the fields $c^a(p)\,$, $c^b(q)$ and
$\bc^c(k)$ and setting the fields to zero.
We obtain
\beq\nome{ST1}
\G_\mu^{(uc)}(p)\,
\G_{\mu}^{(\bc cA)}(k,q,p)
+
\G_\mu^{(uc)}(q)\,
\G_{\mu}^{(\bc cA)}(k,p,q)
+
\G^{(\bc c)}(k)\,
\G^{(vcc)}(k,p,q)=0\,.
\eeq
By taking $p$, $q$ and $k$ at the symmetric point and
using \re{bcc}, \re{uc} and \re{bccA} we obtain
\beq\nome{rovcc}
\r_{vcc}(0)=-\frac{2}{g}
\left[ \frac {k_\mu p_\nu}{p^2} \tilde \G_{\mu\nu}^{(wcA)}(k,q,p)
\right]_{3SP}
\,,
\eeq
where we used $\G^{(vcc)}(p,q,k)|_{3SP}=-1+\r_{vcc}(0)$
and $\G^{(wcA)}(p,q,k)|_{3SP}=\s_{wcA}(0)=-g$.

We discuss now the determination of value of $\r_{4A}(0)$.
{}From the ST identities for three and four vector vertices, one deduces
$$
\left[ p_\mu q_\nu k_\r h_\s
\G_{2;\mu\nu\r\s}^{(4A)}(p,q,k,h)\right]_{4SP}=0\,,
$$
which gives
\beq\nome{ro4A}
\r_{4A}(0) = - \left[ \frac{3}{p^4}
p_\mu q_\nu k_\r h_\s \, \tilde \G^{(4A)}_{2;\mu\nu\r\s}(p,q,k,h)
\right]_{4SP}
\,.
\eeq

\section{Integral equation}
Once the boundary conditions \re{bcirr}, \re{bcs}, \re{rovcc} and
\re{ro4A} are given, the RG equation \re{eveq} allows one to compute
all vertices, at least perturbatively. The nine relevant parameters
are fixed to satisfy the ST identities by definition and this avoids
the fine tuning problem. The couplings $\s_i(\L_0)$ and $\r_i(\L_0)$
can be computed as a function of $g$ and $\L_0/\mu$ and they will be
already fine tuned.

The boundary conditions we use are of the mixed type: the irrelevant
vertices are fixed at $\L=\L_0$, the relevant couplings at $\L=0$.
Moreover, two of the relevant couplings are given in terms of
irrelevant vertices at the physical point $\L=0$. This kind  of
boundary conditions does not allow to solve the RG equation as an
evolution equation, as in the usual case in which one fixes both
irrelevant and relevant vertices at $\L=\L_0$.
In our case the best way to formulate the problem is to convert the RG
equation \re{eveq} into a Volterra integral equation which embodies
the boundary conditions. As we shall show this formulation is also
quite suitable for generating the iterative solution, \ie loop
expansion, and for proving renormalizability. Moreover, it could be
used in principle also for nonperturbative studies.

We introduce the vertices of the functional $I[\Phi;\L]$, defined in
\re{eveq}
\beq\nome{Ivert}
I_{C_1\cdots C_n}(p_1,\cdots p_n;\L)=
-\frac i 2 \int_q M_{BA}(q;\L)\,
\bG_{AB,C_1 \cdots C_n}(-q,q;p_1,\cdots p_n;\L)\,.
\eeq
The functional $I[\Phi;\L]$ can be separated into its relevant and
irrelevant part
$$
I[\Phi;\L]=I_{rel}[\Phi;\L]+I_{irr}[\Phi;\L]\,.
$$
To extract the relevant parameters in $I_{rel}[\Phi;\L]$ we proceed
as in subsect. 3.1. We consider the six vertices in $I[\Phi;\L]$
with non-negative dimension and perform the Lorentz
decomposition and the Taylor type of expansion as done in
subsect.~3.1. For example from the following vertices of $I[\Phi;\L]$
\beq\nome{Ir}
\eqalign{
&
I_{\mu\nu}(p;\L)=
g_{\mu\nu}\,I_L(p;\L)+ t_{\mu\nu}\,I_T(p;\L)\,, \cr
&
I_{\mu\nu}^{(wcA)}(p,q,k;\L)=
g_{\mu\nu}\,I^{(wcA)}(p,q,k;\L)
+\tilde I_{\mu\nu}^{(wcA)}(p,q,k;\L)\,,
\cr&
I_{2;\mu\nu\r\s}^{(4A)}(p,q,k,h;\L)
=
(2g_{\mu\r} g_{\nu\s}\,+ g_{\mu\nu} g_{\r\s}\,)\;
I_2^{(4A)}(p,q,k,h;\L)
+
\tilde I_{2;\mu\nu\r\s}^{(4A)}(p,q,k,h;\L)\,,
}
\eeq
we extract the relevant parameters
\begin{eqnarray*}
&& I_{L}(p;\L)=I_{L}(0;\L)+p^2\partial_{\bp^2}I_{L}(\bp;\L)+\cdots
\equiv
S_{m_A}(\L)+p^2 S_\alpha(\L)+\cdots\,,
\\
&& I_{T}(p;\L)=p^2\partial_{\bp^2}I_{T}(\bp;\L)+\cdots
\equiv
p^2 S_A(\L)+\cdots\,,
\\
&& I^{(wcA)}(p,q,k;\L)=I^{(wcA)}(p,q,k;\L)|_{3SP}+\cdots
\equiv
S_{wcA}(\L)+\cdots\,,
\\
&& I_{2}^{(4A)}(p,q,k,h;\L)=I_2^{(4A)}(p,q,k,h;\L)|_{4SP}+\cdots
\equiv
R_{4A}(\L)+\cdots\,,
\end{eqnarray*}
where the dots stand for irrelevant parts.
The relevant part of the functional $I[\Phi;\L]$ can be again written
in terms of $\Pi_{1,2;rel}$
$$
I_{rel}[\Phi;\L]=\Pi_{1;rel}[\Phi;S_i(\L)]+\Pi_{2;rel}[\Phi;R_i(\L)]
\,.
$$
The integral equations for the relevant and irrelevant parts are then
\beq\nome{int1}
\Pi_{rel}[\Phi;\L]=\Pi_{rel}[\Phi;0]+\int_0^\L \frac{d\l}{\l}
I_{rel}[\Phi;\l]\,,
\eeq
\beq\nome{int2}
\Pi_{irr}[\Phi;\L]=-\int_\L^{\L_0} \frac{d\l}{\l} I_{irr}[\Phi;\l]\,.
\eeq
Finally, by using the form of $\Pi_{rel}[\Phi;0]$, we have
\beq\nome{inteq}
\Pi[\Phi;\L]= S_{int}[\Phi] + \Pi_{2;rel}[\Phi;\r_i(0)]
+  \int_0^\L \, \frac{d\l}{\l}I[\Phi;\l]
-  \int^\infty_0 \, \frac{d\l}{\l} \left\{I[\Phi;\l]-
I_{rel}[\Phi;\l] \right\}
\,,
\eeq
where $S_{int}[\Phi]$ is the interacting part of the BRS action
\re{Stot}. $\Pi_{2;rel}[\Phi;\r_i(0)]$ contains the two additional
relevant couplings fixed by the ST constraint and given by
\beq\nome{r0}
\eqalign{&
\r_{vcc}(0)
=\frac 2 g \int_0^\infty \frac{d\l}{\l} \left[
\frac {k_\mu p_\nu}{p^2} \tilde I^{(wcA)}_{\mu\nu}(k,q,p;\l)
\right]_{3SP}\,,
\cr &
\r_{4A}(0)
=3 \int_0^\infty \frac{d\l}{\l} \left[
\frac{p_\mu q_\nu k_\r h_\s}{p^4} \tilde
I^{(4A)}_{2;\mu\nu\r\s}(p,q,k,h;\l) \right]_{4SP}\,,
}
\eeq
where $\tilde I_{\mu\nu}^{(wcA)}$ and $\tilde I_{2;\mu\nu\r\s}^{(4A)}$
are given in \re{Ir}.

In the last $\l$ integration in \re{inteq} we have sent \UV.
The proof that the subtractions make all integrals finite
in the \UV limit is sketched in subsect.~5.4.
In the next section, where we compute at one loop some vertex functions,
we will see how the subtractions make all integrals finite in the \UV
limit. Only for $\L=0$ the total effective action
$\G[\Phi]=S_2+\Pi[\Phi;0]$ must satisfy the ST identities. This will be
analyzed in the next section at one loop level.

\section{Loop expansion}
The loop expansion is obtained by solving iteratively \re{inteq}.
The starting point is the zero loop order
$\Pi^{(0)}_{1;rel}[\Phi;\s_i(0)]=S_{int}[\Phi]\,.$
This corresponds to
give the relevant vertices in the tree approximation, \ie
\begin{eqnarray*}
&&
\s_{m_A}^{(0)}(\L)=\s_{\alpha}^{(0)}(\L)=
\s_{A}^{(0)}(\L)=\s_{wc}^{(0)}(\L)=0\,,
\\ &&
\s_{3A}^{(0)}(\L)=ig
\,,\;\;\;
\s_{4A}^{(0)}(\L)=-g^2
\,,\;\;\;
\s_{wcA}^{(0)}(\L)=-g\,,
\\ &&
\r_{4A}^{(0)}(\L)=\r_{vcc}^{(0)}(\L)=0\,.
\end{eqnarray*}
At zero loop we have also the auxiliary vertices in $\bG^{(0)}[\Phi;\L]$.
For example we have the vector-vector contributions to order $g^2$
given in fig.~2. They correspond to the usual Feynman diagrams at tree
level with cutoff propagators.

Assume we have computed the cutoff effective action
$\Pi^{(\ell)}[\Phi;\L]$ up to loop $\ell$.
At the next loop this functional can be computed iteratively by the
following steps.
First from eq.~\re{Ivert} we  compute the vertices
$I^{(\ell+1)}_{C_1\cdots C_n}(p_1,\cdots p_n;\l)$.
{}From eq.~\re{Ir} and eq.~\re{r0} we obtain
the vertices $\tilde I_{\mu\nu}^{(wcA)(\ell+1)}(\L)$
and $\tilde I_{2;\mu\nu\r\s}^{(4A)(\ell+1)}(\L)$
and the couplings
$\r_{vcc}^{(\ell+1)}(0)$ and $\r_{4A}^{(\ell+1)}(0)$.
We then obtain the functionals
$I^{(\ell+1)}[\Phi;\L]$, $\,I^{(\ell+1)}_{rel}[\Phi;\L]$ and
$\Pi^{(\ell+1)}_{2;rel}[\Phi;\r_i(0)]$.
Finally from eq.~\re{inteq} we compute
the functional $\Pi^{(\ell+1)}[\Phi;\L]$.

\subsection{One loop vertex functions}
In this subsection we compute to one loop the six vertices
with $\dim \le 4$ which contain the relevant couplings.
We also evaluate to one loop the couplings $\s_i^B$ and $\r_i^B$
of the bare action \re{intac} given by $\s_i(\L_0)$ and
$\r_i(\L_0)$.

\noindent
1) Vector propagator.

\noindent
We have three contributions to the integrand, corresponding to the three usual
Feynman amplitudes
\beq\eqalign{
& I_{\mu\nu}(p;\L)= \;
- \frac{i g^2}{3}
\dL \int_q \frac{K_{\L\L_0}(q)}{q^2}
t^{abba}_{\mu\alpha\alpha\nu}
\cr&
\;\;\; - i g^2
\dL \int_q \frac{K_{\L\L_0}(q,p+q)}{q^2 (p+q)^2}
\left[
V_{\mu\alpha\beta}(p,q,-p-q)
V_{\alpha\beta\nu}(-q,p+q,-p)
+
q_\mu (p+q)_\nu
\right]
\,,
}
\eeq
where $K_{\L\L_0}(q,p+q)=K_{\L\L_0}(q)K_{\L\L_0}(p+q)$ and
$$
V_{\mu\nu\r}(p,q,k)=
g_{\mu\nu}(p-q)_\r
+g_{\nu\r}(q-k)_\mu
+g_{\mu\r}(k-p)_\nu
\,,
$$
\begin{eqnarray*}
t^{a_1 \cdots a_4}_{\mu_1 \cdots \mu_4}
&=&
\left(\eps^{a_1a_2c}\eps^{ca_3a_4}-\eps^{a_1a_4c}\eps^{ca_2a_3}
\right)g_{\mu_1\mu_3}g_{\mu_2\mu_4} +
\left(\eps^{a_1a_3c}\eps^{ca_2a_4}-\eps^{a_1a_4c}\eps^{ca_3a_2}
\right)g_{\mu_1\mu_2}g_{\mu_3\mu_4} \\
&+&
\left(\eps^{a_1a_3c}\eps^{ca_4a_2}-\eps^{a_1a_2c}\eps^{ca_3a_4}
\right)g_{\mu_1\mu_4}g_{\mu_2\mu_3}\,,
\end{eqnarray*}
which are the three and four vector elementary vertices, respectively.
Notice that there is a factor $i$ coming from the $q$-integration.
Thus we get
\beq\nome{vectorI}
I_{L}(p;\L)=
-i g^2 \dL \left[
\int_q \frac{K_{\L\L_0}(q,p+q)}{q^2 (p+q)^2}
(2 q^2+10 pq+3p^2+8\frac{(pq)^2}{p^2})
-6 \int_q \frac{K_{\L\L_0}(q)}{q^2} \right] \,,
\eeq
\beq
I_{T}(p;\L)=
\frac{-ig^2}{3} \dL \left[
\int_q \frac{K_{\L\L_0}(q,p+q)}{q^2 (p+q)^2}
(8 q^2-24 pq+6 p^2-32 \frac{(pq)^2}{p^2})
\right] \,.
\eeq
The relevant couplings for large $\L$ are
\beq
\eqalign{
&
\s^{(1)}_{m_A}(\L) =
\int_0^{\L}\frac{d\l}{\l}I_L(0;\l)
= ig^2 \int_q \frac{K_{0\L}(q)}{q^4} (8\frac{(pq)^2}{p^2}-4 q^2)
= \frac{g^2}{8 \pi^2} \L^2 + {\cal O}(1) \,.
\cr&
\s^{(1)}_{\alpha}(\L)=
\int_0^{\L}\frac{d\l}{\l} \frac{\partial}{\partial \bp^2} I_L(\bp;\l)
={\cal O}(1) \,,
\cr&
\s^{(1)}_{A}(\L)=
\int_0^{\L}\frac{d\l}{\l} \frac{\partial}{\partial \bp^2} I_T(\bp;\l)
=-\frac{10}{3}\frac{g^2}{16 \pi^2} \log (\frac{\L^2}{\mu^2}) + {\cal O}(1) \,.
}
\eeq
{}From \re{int2} the irrelevant parts of the vector propagator are
$$
\Sigma_{L,T}(p;\L)=-\int_\L^{\L_0} \frac{d\l}{\l}
\left[ I_{L,T}(p;\l)-I_{L,T}(0;\l)-p^2
\frac{\partial}{\partial \bp^2} I_{L,T}(\bp;\l)
\right] \,.
$$
Because of the subtractions, we can take the \UV limit. The physical
value is obtained setting $\L=0$ and we have
$$
\Sigma_L(p;0)=
-i g^2 \int_q \left[
\frac{2 q^2+10 pq+3p^2+8\frac{(pq)^2}{p^2}}{q^2 (p+q)^2}
-\frac{4}{q^4}-p^2\frac{\partial}{\partial \bp^2}
\frac{2 q^2+10 \bp q+3\bp^2+8\frac{(\bp q)^2}{\bp^2}}{q^2 (\bp+q)^2}
\right] \,,
$$
$$
\Sigma_T(p;0)=
\frac{-ig^2}{3} \int_q \left[
\frac{8 q^2-24 pq+6 p^2-32 \frac{(pq)^2}{p^2}}{q^2 (p+q)^2}
-p^2\frac{\partial}{\partial \bp^2}
\frac{8 q^2-24 \bp q+6 \bp^2-32 \frac{(\bp q)^2}{\bp^2}}{q^2 (\bp+q)^2}
\right] \,.
$$
In subsect.~5.3 we will show that $\S_L(p;0)=0$, namely the vector propagator
is
transverse, as required by the ST identities.
For $\Sigma_T(p;0)$, for instance, we can write
$$
\frac{\partial}{\partial p^2}\Sigma_T(p;0)
=-\frac{10}{3}\frac{g^2}{16 \pi^2} \log (\frac{p^2}{\mu^2})\,.
$$

\noindent
2) Ghost propagator.

\noindent
The ghost propagator is given in terms of $\Pi^{(wc)}$.
We have one contribution
\beq\nome{Imu}
p_\mu I^{(wc)}(p;\L)=-2 g^2
\dL \left[ \int_q \frac{K_{\L\L_0}(q,p+q)}{q^2 (p+q)^2}
q_\mu \right]  \,.
\eeq
We obtain then the ghost wave function coupling
\beq
\s^{(1)}_{wc}(\L)=
\int_0^{\L}\frac{d\l}{\l} \frac{\partial}{\partial \bp^2}
\left(\bp^2 I^{(wc)}(\bp;\l)\right)
=-i \frac{g^2}{16 \pi^2} \log (\frac{\L^2}{\mu^2}) + {\cal O}(1) \,,
\eeq
for large $\L$. From \re{int2} the irrelevant part of the $w$-$c$ vertex
is, at $\L=0$
$$
p^2\S^{(wc)}(p;0)=-2 g^2
\int_q \left[ \frac{pq}{q^2 (p+q)^2}-p^2
\frac{\partial}{\partial \bp^2} \frac{\bp q}{q^2 (\bp+q)^2}
\right]  \,,
$$
where again we have removed the UV cutoff since the momentum
integration is convergent.

\noindent
3) $w$-$c$-$A$ vertex.

\noindent
In this case the integrand has two contributions
\begin{eqnarray}\nome{Imunu}
\lefteqn{
I^{(wcA)}_{\mu\nu}(p,k,-p-k;\L)= \dL \biggr\{
i g^3 \int_q \frac{K_{\L\L_0}(q,q+k,p-q)}
{q^2 (q+k)^2 (p-q)^2}
}
&& \nonumber \\ && \times
[g_{\mu\nu}(q^2-qk-2qp)-q_\mu k_\nu +
q_\nu (3k_\mu +p_\mu) -p_\nu k_\mu] \biggr\}
\,.
\end{eqnarray}
The coupling $\s_{wcA}(\L)$ for large $\L$ is
\beq
\s^{(1)}_{wcA}(\L)=
\int_0^{\L}\frac{d\l}{\l} I^{(wc A)}(p,k,q;\l)|_{3SP}
=\frac{g^3}{16 \pi^2} \log (\frac{\L^2}{\mu^2}) + {\cal O}(1) \,.
\eeq
This vertex allows one to obtain the coupling $\r_{vcc}(0)$.
Extracting from \re{Imunu} the vertex $\tilde I_{\mu\nu}^{(wca)}$
and inserting it in \re{r0} we find
\beq
\r^{(1)}_{vcc}(0)=
-i g^2 \int_q \frac{p k + 2q p + q k }
{q^2 (q+k)^2 (p-q)^2}\vert_{3SP}
\simeq\; 4.6878\;\frac{g^2}{16\pi^2}\,,
\eeq
where the value is found by numerical integration
over the Feynman parameters.

\noindent
4) $v$-$c$-$c$ vertex.

\noindent
In this case the integrand has one contribution and we have
\beq\nome{Iv2c}
I^{(vcc)}(p,k,-p-k;\L)=- i g^2 \dL \left[
\int_q \frac{K_{\L\L_0}(q,q+k,p-q)}
{q^2 (q+k)^2 (p-q)^2} \;
(pq-q^2) \right]
\,.
\eeq
The relevant coupling for large $\L$ is
\beq
\r^{(1)}_{vcc}(\L)=
\int_0^{\L}\frac{d\l}{\l} I^{(vcc)}(\bp;\l)
=\frac{g^2}{16 \pi^2} \log (\frac{\L^2}{\mu^2}) + {\cal O}(1) \,.
\eeq

\noindent
5) Three vector vertex.

\noindent
The integrand for the three vector vertex is too long
and we give only the result for the coupling for large $\L$
\beq
\s^{(1)}_{3A}(\L)=
i\frac 4 3 \frac{g^3}{16 \pi^2} \log (\frac{\L^2}{\mu^2}) + {\cal O}(1) \,.
\eeq

\noindent
6) Four vector vertex.

\noindent
Also for this vertex the integrand is too long and we give only the
values for the two couplings for large $\L$
$$
\s^{(1)}_{4A}(\L)=
\frac 2 3 \frac{g^4}{16 \pi^2} \log (\frac{\L^2}{\mu^2}) + {\cal O}(1) \,,
$$
$$
\r^{(1)}_{4A}(\L)=
-10 \frac{g^4}{16 \pi^2} \log (\frac{\L^2}{\mu^2}) + {\cal O}(1) \,.
$$
{}From this vertex we obtain the coupling $\r_{4A}(0)$ given in
\re{r0}. From a numerical integration over the Feynman parameters
we find
\beq
\r^{(1)}_{4A}(0)\simeq\; 1.2329\;\frac{g^2}{16\pi^2}\,.
\eeq

\subsection{One loop beta function}

We discuss in the present formulation the role of the subtraction
point $\mu$ and deduce the beta function (see also ref.~\cite{H} for the
scalar case).
In this subsection we take the physical values $\L=0$ and \UV.
Denote by $\G_{nA,mwc}(g,\mu)$
the vertices with $n$ vector fields and $m$ pairs $w_\mu\, c$, which
satisfy the physical conditions \re{bcs} with couplings $\s_i$  and $\r_i$
defined as in subsect.~3.1 at the scale $\mu$.
Therefore $\mu$ is the only dimensional parameter,
apart for the external momenta.
However physical measurements should not depend on the specific
value of $\mu$.
Consider the vertices $\G_{nA,mwc}(g',\mu')$ with coupling $g'$ at a
new scale $\mu'$.
The request that the two sets of vertices  $\G_{nA,mwc}(g,\mu)$
and $\G_{nA,mwc}(g',\mu')$ describe the same theory
implies that the corresponding effective actions
$\G[\Phi;g,\mu]$  and $\G[\Phi';g',\mu']$ are equal, where the
fields in $\Phi$ and $\Phi'$ are related by a rescaling,
$A'_\mu(p)=\sqrt{Z_A}A_\mu(p)$ and $w_\mu'(q)c'(p)=Z_c w_\mu(q) c(p)$.
This implies that the two sets of vertices are related by
$$
\G_{nA,mwc}(g',\mu')=Z_A^{-n/2} Z_c^{-m} \G_{nA,mwc}(g,\mu) \,.
$$
To obtain $Z_A$ we use \re{propp}-\re{st}, which give
$$
Z_A=1-\frac{\partial}{\partial p^2}\S_T(p;g,\mu)|_{p^2=\mu'^2}
\equiv 1-f_A(\mu';g,\mu) \,.
$$
Similarly from \re{wc} we find
$$
Z_c=1+i\frac{\partial}{\partial p^2}\left[p^2 \S^{(wc)}(p;g,\mu)
\right]|_{p^2=\mu'^2}
\equiv 1+f_c(\mu';g,\mu)  \,.
$$
We can now give the expression of the beta function.
By using the physical condition \re{wcA} for the $w$-$c$-$A$ vertex
and the relation
$\G^{(wcA)}_{\alpha\beta}(g',\mu')=Z_A^{-1/2}Z^{-1}_c
\G^{(wcA)}_{\alpha\beta}(g,\mu)$
we obtain the renormalization group relation
$$
g'= \frac{g-f_g(\mu';g,\mu)}
{\left( 1+f_c(\mu';g,\mu) \right)\left( 1-f_A(\mu';g,\mu) \right)^{1/2}}
\,,
$$
where $f_g(\mu';g,\mu)=\S^{(wca)}(p,q,k;g,\mu)|_{3SP'}$.
The beta function is obtained by considering an infinitesimal
scale change and is given by
$$
\beta(g)=\mu'\partder{\mu'}
\left\{-f_g(\mu';g,\mu)-g\,f_c(\mu';g,\mu)+\half g\, f_A(\mu';g,\mu)
\right\}|_{\mu'=\mu}\,.
$$
The three dimensionless quantities $f_i(\mu';g,\mu)$ are functions of $g$
and the ratio $\mu'/\mu$, thus the beta function depends only on $g$.
{}From the above results we find, at one loop order,
$$
f_A=-\frac{10}{3}\frac{g^2}{16\pi^2}\log(\frac{\mu'^2}{\mu^2})
\,,\;\;\;\;
f_c=\frac{g^2}{16\pi^2}\log(\frac{\mu'^2}{\mu^2})
\,,\;\;\;\;
f_g=\frac{g^3}{16\pi^2}\log(\frac{\mu'^2}{\mu^2})
\,.
$$
Thus
$$
\beta^{(1)}(g)=
-\frac {11}{3} \frac {1}{16\pi^2}g^3 \mu'\partder{\mu'} \ln\frac{\mu'^2}{\mu^2}
=-\frac {22}{3}\frac{g^3}{16\pi^2} \,,
$$
which is the usual one loop result.

Since there is only one coupling, the beta function can be defined
also from the three and four vector vertex. One finds
$$
\beta(g)=\mu'\partder{\mu'}
\left\{-if_{3A}(\mu';g,\mu)+\frac 3 2 g\, f_A(\mu';g,\mu)
\right\}|_{\mu'=\mu}
$$
and
$$
\beta(g)=\mu'\partder{\mu'}
\left\{-\frac{1}{2g} f_{4A}(\mu';g,\mu)+ g\, f_A(\mu';g,\mu)
\right\}|_{\mu'=\mu}\,,
$$
where $f_{3A}(\mu';g,\mu)=\G^{(3A)}(p_i;g,\mu)|_{3SP'}$ and
$f_{4A}(\mu';g,\mu)=\G^{(4A)}_1(p_i;g,\mu)|_{4SP'}$. At one loop they
are given by
$$
f_{3A}=i\frac{4}{3}\frac{g^3}{16\pi^2}\log(\frac{\mu'^2}{\mu^2})
\,,\;\;\;\;
f_{4A}=\frac{2}{3}\frac{g^4}{16\pi^2}\log(\frac{\mu'^2}{\mu^2})
\,.
$$
We have then that these definitions give the same one loop
value for the beta function.

\subsection{One loop ST identities}
In this subsection we show how two simple ST identities are satisfied
at one loop order in the physical limit. Thus in this subsection we set $\L=0$
and we will take \UV.
The first one is the vector propagator transversality and is obtained from
\re{ST} by derivating with respect to $c^a(p)$ and $A_\mu^b(q)$. We get
$$
\G_\mu^{(uc)}(p) \left[ p_\mu p_\nu + \G_{\mu\nu}(p)\right]=0 \,,
$$
which implies $\Sigma_L(p)=0$.
To show this at one loop, we consider first the unsubtracted integral
$$
\Sigma'_L(p)=-\int_0^{\L_0} \frac{d\l}{\l} I_L(p;\l) \,.
$$
{}From \re{vectorI}, after some algebra, we find
$$
\Sigma'_L(p)=-i \int_q \frac{K_{0\L_0}(q)}{q^2}
\biggr[\frac{ 4pq}{p^2} \left( K_{0\L_0}(q+p)-K_{0\L_0}(q-p) \right)
+6\left( K_{0\L_0}(q+p)-K_{0\L_0}(q) \right) \biggr] \,.
$$
Due to the difference of the two cutoff functions we have that $q^2$
is forced into the region $q^2 \sim \L_0^2$.
By taking for instance an exponential UV cutoff one has
$$
K_{0\L_0}(q+p)- K_{0\L_0}(q-p)=-4\frac{p\cdot q}{\L_0^2}
\left\{1-\frac{p^2}{\L_0^2}+\frac 2 {3}
\frac{(p\cdot q)^2}{\L_0^4} +\cdots\right\}
\, e^{-{q^2}/{\L_0^2}}
$$
and obtains
$$
\Sigma'_L(p) = a\L_0^2 + b p^2 + {\cal O} (\frac{p^2}{\L^2_0})\,,
$$
where $a$ and $b$ are numerical constants.
In this calculation the effect of the non invariant regularization
is clear.
The divergent integral with the cutoff functions gives a surface
term which destroys the transversality of the propagator.
However, the longitudinal contributions are of relevant type, thus they
are cancelled by imposing the boundary conditions and taking the limit \UV.
One finds
$$
\Sigma_L^{(1)}(p)=\Sigma'_L(p)-\Sigma'_L(0)
-p^2 \partial_{\bp^2} \Sigma'_L(\bp)|_{\bp^2=\mu^2}=
{\cal O} (\frac{p^2,\mu^2}{\L^2_0}) \to 0\,.
$$
Notice that for $\L\neq 0$ the longitudinal part of the photon propagator
is different from zero both in the relevant ($\s_{m_A}(\L)$ and
$\s_\alpha(\L)$) and irrelevant ($\S_L(p;\L)$) parts.

We now verify the identity \re{ST1}. At one loop order it becomes
\begin{eqnarray*}
&&
\frac 1 g \int_0^{\L_0} \frac{d\l}{\l}
\biggr[ p_\nu k_\mu \left( I_{\mu\nu}^{(wcA)}(k,h,p;\l)
-g_{\mu\nu}I_{rel}^{(wcA)}(\mu^2;\l) \right) +
\;\;\;\;h\leftrightarrow p \;\; \biggr]
\\ &&
-i \int_0^{\L_0} \frac{d\l}{\l}\biggr[ k p \left( I^{(wc)}(p;\l)-
I_{rel}^{(wc)}(\mu^2;\l)\right)+\;\;\; p\rightarrow h \;\;\;+
\;\;\;p \rightarrow k \biggr]
\\ && +
k^2 \int_0^{\L_0} \frac{d\l}{\l} \left( I^{(vcc)}(k,p,h;\l)-
I_{rel}^{(vcc)}(\mu^2;\l) \right)
-k^2 \r_{vcc}(0) \to 0 \;\;\;\;\; \mbox{for} \;\L_0\to \infty\,.
\end{eqnarray*}
As above we keep the UV cutoff $\L_0$, so that we can consider separately
the integrands $I$ and their subtractions $I_{rel}$.
{}From \re{Imu}, \re{Imunu} and \re{Iv2c}
the unsubtracted contribution can be cast in the form
$$
ig^2k_\mu\biggr\{\int_q k_\mu\frac{K_{0\L_0}(q,q+k)}{q^2(q+k)^2}
\left[K_{0\L_0}(q-h)-1\right]\;\;+\;\;
(k\rightarrow  h,\;h\rightarrow  k)\;\;+\;\;
(k\rightarrow  p,\; h\rightarrow  h)\biggr\}\,.
$$
As in the previous case, the factor $(K_{0\L_0}-1)$ forces the
integration momentum to be of the order $q^2 \sim \L_0^2$ so that
the integral vanishes for \UV.
It is easy to verify that also the remaining terms vanish for \UV.

\subsection{Perturbative renormalizability}
Perturbative renormalization is essentially based on power counting.
To prove that the theory is perturbatively renormalizable one has to
show  that the integral equations give a finite result in the limit \UV.
This can be done perturbatively by iterating eqs.~\re{int1} and \re{int2}.
In order to see that the integrations
over $\l$ are convergent, we have to estimate the behaviour of the integrands
for large $\l$.
The analysis can be simplified, following Polchinski, by introducing the norm
$$
|\G(n_A,n_{\bc c},n_u,n_v)|_\L\,
 \equiv \Maxlp |\G_{C_1\cdots C_n}(p_1,\cdots p_n;\L)|\,,
$$
which allows us to ignore the momentum dependence.
Since the $\L$-dependence is fixed only by the
number of fields, to simplify the notation, we have indicated
in the vertices only the numbers $n_A$ of vectors, $n_{\bc c}$
of ghost-antighost pairs, $n_u$ and $n_v$ of $u$ and $v$ sources.
We deduce perturbative renormalizability by proving that the
$\L$ dependence of this norm is given by power counting.
We show this by induction on the number of loops.
Namely, at loop $\ell$, we assume for large $\L$ the
following behaviour
\beq\nome{pwc}
|\G^{(\ell)}(n_A,n_{\bc c},n_u,n_v)|_\L\,
=\,{\cal O}(\L^{4-n_A-2n_{\bc c}-2n_u-2n_v})\,,
\eeq
which is satisfied at $\ell=0$.
We neglect for simplicity all possible $\ell$-dependent powers of
$\log\frac{\L}{\mu}$.

We now proceed by iteration and prove that the behaviours \re{pwc} are
reproduced at the loop $\ell+1$.
We follow the method of ref.~\cite{BDM,QED}.
First of all one notices that the norm of the auxiliary vertices have
the same behaviours as the corresponding vertices, as can be
seen from their definition. We have then
\beq
|\bG^{(\ell)}_{AB}(-q,q;n_A,n_{\bc c},n_u,n_v;\L)|_\L=
{\cal O}(\L^{2-n_A-2n_{\bc c}-2n_u-2n_v})\,.
\eeq
{}From this behaviour one obtains the bound for the integrands \re{Ivert}
of the vertices at loop $\ell+1$
\beq
|I^{(\ell+1)}(n_A,n_{\bc c},n_u,n_v)|_\L\,
=\,{\cal O}(\L^{4-n_A-2n_{\bc c}-2n_u-2n_v})\,,
\eeq
where one uses the fact that $M_{AB}$ gives a factor ${\L}^{-2}$.
Moreover, the derivative of the integrands $I^{(\ell+1)}$
with respect to the external momenta reduces the powers of $\L$
\beq\nome{der}
|\partial_p^m I^{(\ell+1)}(n_A,n_{\bc c},n_u,n_v)|_\L\,
=\,{\cal O}(\L^{4-n_A-2n_{\bc c}-2n_u-2n_v-m})\,.
\eeq
The proof of this is given for instance in ref.~\cite{BDM}.

Finally, from these behaviours to obtain the vertices at loop $\ell+1$.
For the relevant couplings the integrand in \re{int1}
grows with $\l$ and the result is dominated by the upper limit $\L$.
This reproduces immediately \re{pwc}.

For the irrelevant vertices, which are given by \re{int2}, the
$\l$-integration goes up to infinity.
We treat separately the irrelevant vertices with negative and
non-negative dimension. For the first ones
the integration over $\l$ is convergent and the integral
is dominated by the lower limit, thus reproducing
at loop $\ell+1$ the assumption \re{pwc}. For the second ones
due to the subtractions the integrand can be expressed as
derivative with respect to the external momenta for
which we can use \re{der}.
Consider for instance the case of $\S^{(\ell+1)}_L(p;\L)$
We have
\beq
|I^{(\ell+1)}_L(p;\l)-I^{(\ell+1)}_{rel\,L}(p;\l)|_\l\sim p^4|
\partial^4 I^{(\ell+1)}_L(p;\l)|_\l
=p^4{\cal O}(\l^{-2})
\eeq
for $p^2<\l^2$.
Then by inserting the behaviour \re{der} in \re{int2} we have that
the $\l$-integration is convergent and dominated by the lower limit $\L$.
Thus the assumption \re{pwc} is recovered.

\section{Functional identity}
In this paper we have assumed that the interaction part of the cutoff
effective action depends only on the combination of fields $w_\mu(p)$,
namely we have the following functional identity
\beq\nome{FI}
-igp_\mu\frac{\delta\Pi[\Phi;\L]}{\delta u_\mu^a(p)}=
\frac{\delta\Pi[\Phi;\L]}{\delta \bc^a(p)}\,.
\eeq
Since $w_\mu$ has mass dimension $2$ we have reduced the number of
vertices with non-negative dimension and consequently the number of
relevant couplings. Moreover, this identity simplifies the ST identities.

In this section we will prove that, by a proper choice of the boundary
conditions of the evolution equation, this identity is maintained at
all loops.

If we do not assume \re{FI} the fields $\bc$ and $u_\mu$ enter
independently in the effective action.
Then instead of the two monomials containing $w_\mu$ in \re{7s}
we would have four relevant couplings associated to the monomials
$$
\bc \cdot c \,, \;\;\;u_\mu \cdot c \,,\;\;\;
(\partial_\mu \bc) \cdot (c \wedge A_\mu) \,, \;\;\;
u_\mu \cdot (c \wedge A_\mu)\,.
$$
The four couplings associated with these monomials are defined by
\begin{eqnarray*}
&&\Pi^{(\bc c)}(p;\L)=
i p^2 (\s_{\bc c}(\L) +\S^{(\bc c)}(p;\L))\,,
\\
&&\Pi_\mu^{(uc)}(p;\L)=
p_\mu (\s_{uc}(\L)+\S^{(uc)}(p;\L)) \,,
\\
&&\G_{\mu}^{(\bc cA)}(p,q;\L)=p_\mu[\s_{\bc cA}+
\S^{(\bc cA)}(p,q;\L)]+
q_\mu \tilde\G^{(\bc cA)}(p,q;\L)\,,
\\
&&\G_{\mu\nu}^{(ucA)}(p,q;\L)=
g_{\mu\nu}[\s_{ucA}(\L) + \S^{(ucA)}(p,q;\L)]+
\tilde \G_{\mu\nu}^{(ucA)}(p,q;\L)\,,
\end{eqnarray*}
where the normalization conditions are
$$
\partial_{p^2} \left( p^2 \S^{(\bc c)}(p;\L) \right)|_{p^2=\mu^2}=
\partial_{p^2} \left( p^2 \S^{(uc)}(p;\L) \right)|_{p^2=\mu^2}=0\,,
$$
$$
\S^{(\bc cA)}(p,q;\L)|_{3SP}=
\S^{(ucA)}(p,q;\L)|_{3SP}=0 \,.
$$
The evolution equations and their integral solutions are completely analogous
to the previous case.

The field $\bc$ enters in four other monomials of dimension four
$$
(\partial_\mu c) \cdot (\bc \wedge A_\mu) \,, \;\;\;
(\bc \wedge c)\cdot (A_\mu \wedge A_\mu)\,, \;\;\;
(\bc \cdot A_\mu) (c \cdot A_\mu)\,, \;\;\;
(\bc \wedge c)\cdot (\bc \wedge c)\,.
$$
If the functional identity holds,
whenever there is a $\bc$ field in a vertex, we can extract a power
of its momentum and we are left with an irrelevant vertex.
We can therefore assume that there are no relevant couplings
associated with these monomials (notice that for this reason we have assumed
that there is only one relevant coupling in $\G_{\mu}^{(\bc cA)}$).
The corresponding vertices are then fixed to vanish at $\L=\L_0$.

Since the Lorentz decomposition and the normalization conditions for the
vertices $w$-$c$ and $w$-$c$-$A$ in \re{uc} and \re{wcA} are analogous to
the one for the vertices $u$-$c$ and $u$-$c$-$A$ given above, we select the
boundary conditions
$$
\s_{u c}(0)=\frac 1 g \s_{w c}(0)=0\,,\;\;\;\;\;
\s_{u cA}(0)=\frac 1 g \s_{w cA}(0)=-1\,.
$$
This corresponds to fix these two couplings as they appear in the total action
\re{Stot}.
We now show that it is possible to fix the boundary conditions on the
parameters $\s_{\bc c}$ and $\s_{\bc cA}$ in such a way that
the identity \re{FI} is valid at any loop and for any $\L$.
At the tree level this requirement gives
$$
\s_{\bc c}^{(0)}(0)=0\,,\;\;\;\;\;
\s_{\bc cA}^{(0)}(0)=ig\,.
$$
Let us suppose that the identity is true for the cutoff effective action
$\Pi^{(\ell)}$ at loop $\ell$.
The identity then holds for the auxiliary vertices $\bG^{(\ell)}$ and,
from \re{Ivert}, also for the integrands $I^{(\ell+1)}$.
Then the validity of \re{FI} at loop $\ell+1$ depends only on the boundary
conditions for $\s_{\bc c}^{(\ell+1)}$ and $\s_{\bc cA}^{(\ell+1)}$.
In the following, in order to simplify the notation, in
all the integrands and vertices we do not write the loop order which is
understood to be $\ell+1$.

\noindent
1) Ghost propagator.

The integral equations give ($\bp^2=\mu^2$)
$$
\Pi^{(\bc c)}(p;\L)=ip^2 \s^{(\ell+1)}_{\bc c}(0)+
ip^2\int_0^{\L_0}\,\frac{d\l}{\l}
\partial_{\bp^2} \bp^2 I^{(\bc c)}(\bp;\l)
-ip^2 \int_\L^{\L_0}\,\frac{d\l}{\l}
I^{(\bc c)}(p;\l)
$$
and
$$
p_\mu\Pi_\mu^{(uc)}(p;\L)=p^2\int_0^{\L_0}\,\frac{d\l}{\l}
\partial_{\bp^2} (\bp^2 I^{(u c) }(\bp;\l))
-p^2 \int_\L^{\L_0}\,\frac{d\l}{\l}
I^{(u c) }(p;\l) \,.
$$
{}From the inductive hypothesis
$$
g I^{(u c) }(p;\l)=I^{(\bc c) }(p;\l)\,,
$$
we have that the identity is satisfied at
any loop and for any $\L$, by imposing $\s_{\bc c}^{(\ell+1)}(0)=0$, which
implies to all loop order
$$
\s_{\bc c}(0)=0\,.
$$

\noindent
2) Vector-ghost vertex.

The integral equations give
$$
\G_\mu^{(\bc cA)}(p,q;\L)= p_\mu\s_{\bc cA}^{(\ell+1)}(0)
+p_\mu \int_0^{\L_0}\,\frac{d\l}{\l}I^{(\bc c A)}(p,q;\l)\vert_{3SP}
-\int_\L^{\L_0}\,\frac{d\l}{\l}
I_\mu^{(\bc c A)}(p,q;\l)
$$
and
$$
\G_{\nu\mu}^{(u cA)}(p,q;\L)= -g_{\mu\nu}
+g_{\mu\nu}\int_0^{\L_0}\,\frac{d\l}{\l}I^{(uc A)}(p,q;\l)\vert_{3SP}
-\int_\L^{\L_0}\,\frac{d\l}{\l}
I_{\nu\mu}^{(uc A)}(p,q;\l)
\,.
$$
The inductive hypothesis
$$
-ig p_\nu I_{\nu\mu}^{(u c A)}(p,q;\l)=I_\mu^{(\bc c A)}(p,q;\l)\,,
$$
at $3SP$ gives
$$
-ig \left[I^{(u c A)}(p,q;\l)+\frac{2}{3p^2} p_\mu Q_\nu
\tilde I_{\mu\nu}^{(ucA)}(p,q;\l)\right]_{3SP}
=I^{(\bc c A)}(p,q;\l)\vert_{3SP}\,,
$$
where $Q_\nu=2p_\nu+q_\nu$ is vector orthogonal to $q_\nu$ at $3SP$.
Using these two results one finds that in this case the identity \re{FI}
is valid for any $\L$ if we fix
$$
\s^{(\ell+1)}_{\bc c A}(0)=ig\left[\frac{2}{3p^2} Q_\alpha p_\beta
\int_0^{\L_0}\,\frac{d\l}{\l}
\tilde I^{(u c A)}_{\beta\alpha}(p,q;\l)\right]_{3SP}\,.
$$
To all loops we have then
$$
\s_{\bc c A}(0)=ig \left[1-\frac{2}{3p^2}  p_\beta
\tilde \G^{(u c A)}_{\beta\alpha}(p,q;0)Q_\alpha|_{3SP}\right]\,.
$$

Notice that this condition implies
$$
-ig\s_{u c A}(\L_0)=
\s_{\bc c A}(\L_0)\,,
$$
namely, in the usual language, no counterterms violating \re{FI} are
generated by the evolution equation.

\section{Conclusions}
In this paper we have analyzed a RG formulation of a non-Abelian
gauge theory which avoids the usual fine tuning problem
\cite{B,romani}. This problem arises when one treats the RG equations
as evolution equations and starts by giving the relevant couplings at
the UV scale $\L_0$ (bare parameters). These various parameters have
to be fine tuned to depend on a single parameter, the coupling $g$ at
the normalization point $\mu$,
in such a way that the ST identities are satisfied by the
computed physical effective action. Our procedure avoids the fine
tuning problem simply by fixing the relevant parameters at the
physical point $\L=0$. We have shown that the formulation so obtained is quite
practical and deduced the usual loop expansion of the renormalized
vertex functions in terms of the physical coupling $g$ at $\mu$.
The calculations do not seem more complex than the ones done in
dimensional regularization.
The cancellations of UV divergences in the Feynman graphs is provided
by subtractions which are systematically generated by implementing the
boundary conditions. These cancellations are very similar to the
ones obtained by the Bogoliubov $R$-operation \cite{Bog}.
We described very briefly the proof of perturbative renormalizability.
This proof is simply based on power counting and does not involve any
new feature with respect to the one for the scalar case and QED.

We have performed detailed calculations only to one loop.
We computed the six vertices with non-negative dimension
which contain the relevant couplings.
{}From them we obtained for large $\L$ the nine relevant couplings $\s_i(\L)$
and
$\r_i(\L)$, which are used to compute the usual one
loop beta function and are the coefficients of the UV bare action
\re{intac}.

One then has to show that the complete (physical) effective action
indeed satisfies ST identities.
In the QED case we have shown \cite{QED} to all perturbative orders
that the complete effective action computed by this method does indeed
satisfy the Ward identities.
This implies that, in the RG flow, all couplings
and vertices are constrained for every $\L$ by the symmetry.
In the YM case we have verified the ST identities only to one loop
and for two vertices. A general proof of ST identities,
based on the results of the recent work \cite{B} by Becchi, is under
study.

Our analysis has been perturbative. However, contrary to dimensional
regularization, the basic integral equations in \re{inteq} and \re{r0}
are in principle nonperturbative. The only possible perturbative
element at the basis of \re{inteq} is simply the separation of
$\G[\Phi;\L]$ in relevant and irrelevant parts, which is based on
na\^{\i}ve power counting. This should be correct for an
asympotically free theory. A next natural step in this formulation is
the analysis of anomalies and chiral gauge theories. Since we work in
four dimensions, the presence of $\g_5$ does not lead to any direct
complication. This analysis is left to a separate publication
\cite{anom}.

\vspace{3mm}\noindent{\large\bf Acknowledgements}

We have benefited greatly from discussions with C. Becchi, G. Bottazzi
and M. Tonin.

\eject

\eject
\newpage
\begin{figcap}

\item
The circles represent vertex functions with the IR cutoff $\L$.
Internal lines involve only the fields $\phi_A=(A_\mu,c,\bar c )$,
while the external lines could involve also the sources $u_\mu$ and $v$.
The crosses represent the derivative with respect to $\L$ of the internal
propagators. Integration over $q$ in the loop is understood.

\item
a) Graphical representation of the contribution to
$\bG^{(0)b_1b_2aa'}_{\nu_1\nu_2,\mu\mu'}(q_1,q_2;p,p';\L)$ obtained
by expanding \re{bG} to second order in $\G^{int}$.

b) Graphical representation of the contribution to
$\bG^{(0)b_1b_2aa'}_{\nu_1\nu_2,\mu\mu'}(q_1,q_2;p,p';\L)$ obtained
by expanding \re{bG} to first order in $\G^{int}$.

c) Graphical representation of the contribution to
$\bG^{(0)b_1b_2aa'}_{\mu\mu'}(q_1,q_2;p,p';\L)$ obtained
by expanding \re{bG} to second order in $\G^{int}$.

\end{figcap}

\end{document}